\documentclass[12pt,eqno,epsf]{article}
\usepackage{amsmath,amssymb,graphicx,slashbox}
\numberwithin{equation}{section}

\def\mydate{February 3, 2011}
\def\ignore#1{{}}

\newcounter{sxn}

\newcounter{axn}

\date{}

\newdimen\mybaselineskip
\renewcommand{\thefootnote}{\arabic{footnote}}

\newcommand{\beeq}{\begin{equation}}
\newcommand{\eneq}{\end{equation}}
\newcommand{\beqn}{\begin{eqnarray}}
\newcommand{\eeqn}{\end{eqnarray}}

\newcommand{\alp}{\alpha}
\newcommand{\bt}{\beta}
\newcommand{\gm}{\gamma}
\newcommand{\Gm}{\Gamma}
\newcommand{\dlt}{\delta}

\newcommand{\ep}{\epsilon}
\newcommand{\zt}{\zeta}
\newcommand{\vep}{\varepsilon}
\newcommand{\tht}{\theta}

\newcommand{\kp}{\kappa}
\newcommand{\lmd}{\lambda}

\newcommand{\sgm}{\sigma}

\newcommand{\omg}{\omega}

\newcommand{\be}{\begin{equation}}
\newcommand{\ee}{\end{equation}}
\newcommand{\bea}{\begin{eqnarray}}
\newcommand{\eea}{\end{eqnarray}}
\newcommand{\eql}{\!\!\!&=\!\!\!&}

\newcommand{\sma}{\!\!\!&\simeq\!\!\!&}
\newcommand{\defa}{\!\!\!&\equiv\!\!\!&}

\newcommand{\simgt}{\stackrel{>}{{}_\sim}}
\newcommand{\simlt}{\stackrel{<}{{}_\sim}}

\newcommand{\tl}[1]{\tilde{#1}}
\newcommand{\bdm}[1]{{\mbox{\boldmath $#1$}}}
\newcommand{\tr}{{\rm tr}}

\newcommand{\diag}{{\rm diag}}
\newcommand{\der}{\partial}
\newcommand{\dr}{\!\!d}
\newcommand{\hc}{{\rm h.c.}}
\newcommand{\ie}{{\it i.e.}}

\newcommand{\brkt}[1]{\left( #1 \right)}
\newcommand{\brc}[1]{\left\{ #1 \right\}}
\newcommand{\sbk}[1]{\left[ #1 \right]}
\newcommand{\abs}[1]{\left| #1 \right|}

\renewcommand{\Re}{{\rm Re}\,}
\renewcommand{\Im}{{\rm Im}\,}

\newcommand{\cD}{{\cal D}}

\newcommand{\cI}{{\cal I}}
\newcommand{\cK}{{\cal K}}
\newcommand{\cL}{{\cal L}}

\newcommand{\cO}{{\cal O}}

\newcommand{\thw}{\tht_W}
\newcommand{\bthw}{\vartheta_W}

\newcommand{\cph}{c_\phi}
\newcommand{\sph}{s_\phi}

\newcommand{\suL}{SU(2)_{\rm L}}
\newcommand{\suR}{SU(2)_{\rm R}}
\newcommand{\uy}{U(1)_Y}
\newcommand{\uem}{U(1)_{\rm EM}}
\newcommand{\mKK}{m_{\rm KK}}

\newcommand{\thH}{\theta_{\rm H}}

\begin{document}
\thispagestyle{empty}

\baselineskip=12pt

{\small \noindent \mydate    
\hfill }

{\small \noindent \hfill  KEK-TH-1352}

\baselineskip=35pt plus 1pt minus 1pt

\vskip 1.5cm

\begin{center}
{\Large \bf Modulus stabilization and IR-brane kinetic terms}\\
{\Large \bf in gauge-Higgs unification}\\

\vspace{1.5cm}
\baselineskip=20pt plus 1pt minus 1pt

\normalsize

{\bf Nobuhito\ Maru},${}^1\!${\def\thefootnote{\fnsymbol{footnote}}
\footnote[1]{\tt e-mail address: maru@phys.chuo-u.ac.jp}} 
{\bf and 
Yutaka\ Sakamura}${}^{2,3}\!${\def\thefootnote{\fnsymbol{footnote}}
\footnote[2]{\tt e-mail address: sakamura@post.kek.jp}}

\vspace{.3cm}
${}^1${\small \it Department of Physics, Chuo University, 
Tokyo, 112-8551, Japan} \\ \vspace{3mm}
${}^2${\small \it KEK Theory Center, Institute of Particle and Nuclear Studies, 
KEK, \\ Tsukuba, Ibaraki 305-0801, Japan} \\ \vspace{3mm}
${}^3${\small \it Department of Particles and Nuclear Physics, \\
The Graduate University for Advanced Studies (Sokendai), \\
Tsukuba, Ibaraki 305-0801, Japan} 
\end{center}

\vskip 1.0cm
\baselineskip=20pt plus 1pt minus 1pt

\begin{abstract}
We discuss the modulus stabilization by the Casimir energy and 
various effects of IR-brane kinetic terms for the gauge fields 
in a gauge-Higgs unification model in the warped spacetime, 
where the Wilson line phase~$\thH$ is determined as $\thH=\pi/2$.  
We find that the brane kinetic terms with $\cO(1)$ coefficients 
are necessary for the modulus stabilization. 
On the other hand, large brane kinetic terms can deviate 
4D gauge couplings from the standard model values and 
also cause too light Kaluza-Klein (KK) modes. 
In the parameter region that ensures the modulus stabilization, 
the KK gluon appears below 1~TeV, 
which marginally satisfies the experimental bound. 
The allowed parameter region will be enlarged in a model
where a smaller value of $\thH$ is realized.  
\end{abstract}


\newpage

\section{Introduction}
The gauge-Higgs unification scenario is an interesting candidate for 
the physics beyond the standard model, which was originally proposed 
in Refs.~\cite{Fairlie:1979at,Hosotani:1983xw} 
and revived by Refs.~\cite{Hatanaka:1998yp,Pomarol:1998sd} 
as a solution to the hierarchy problem. 
In this class of models, the Higgs mass is protected 
against large radiative corrections 
thanks to a higher-dimensional gauge symmetry~\cite{Antoniadis:2001cv}. 
Since the Higgs field is identified with an extra-dimensional component 
of a higher-dimensional gauge field, this class of models do not require 
any elementary scalar fields, which often cause a hierarchy 
problem due to large radiative corrections to their masses. 

The gauge-Higgs unification has been first investigated 
in the flat spacetime~\cite{Csaki:2002ur,Hall:2001zb}. 
These models have common features that 
the physical Higgs boson and the Kaluza-Klein (KK) excitation modes 
become too light to satisfy the experimental bounds 
unless the Wilson line phase~$\thH$ along the extra dimension 
takes a very small value. 
Besides, the large top quark mass is not realized 
in simple models although there is an elaborate way to realize it.\footnote{
In Ref.~\cite{Cacciapaglia:2005da}, the top quark mass is realized by using 
large Clebsch-Gordan coefficients in higher-dimensional representations 
of the matter multiplets.
}
These difficulties are easily solved 
in the Randall-Sundrum warped spacetime~\cite{Randall:1999ee}. 
The Higgs and KK masses are enhanced by 
a logarithm of the large warp factor~\cite{Hosotani:2005nz}, 
and the top quark mass can easily be realized 
only by the localization of the mode functions 
in the extra dimension~\cite{Agashe:2004rs}. 
Furthermore, the gauge-Higgs unification 
in the warped spacetime has phenomenologically interesting 
features~\cite{Agashe:2004rs}-\cite{Hosotani:2009jk}. 
Hence, we will focus on the Randall-Sundrum spacetime 
as a background geometry in this paper. 

When we discuss extra-dimensional models, the stabilization mechanism 
for the size of the extra dimension, which is often called the modulus 
or the radion, must be considered. 
One of the simplest mechanisms 
for the modulus stabilization is proposed in Ref.~\cite{Goldberger:1999uk}. 
A five-dimensional (5D) bulk scalar field plays an essential role 
for the stabilization in this mechanism. 
Thus it spoils one of the virtues of the gauge-Higgs unification, \ie, 
no need to introduce an elementary scalar field. 
There is another way for the modulus stabilization by using 
the Casimir energy of the bulk fields. 
The stability by the Casimir energy has been discussed in many 
papers~\cite{Fabinger:2000jd}-\cite{Ponton:2001hq},  
and it has been shown that the bulk gauge field can provide 
a significant contribution to the effective potential~\cite{Garriga:2002vf}. 
Thus this mechanism is expected to be feasible 
in the gauge-Higgs unification scenario 
because the bulk gauge fields are essential ingredients. 
Besides, this mechanism does not need any elementary scalar fields. 
Therefore it is an intriguing subject to discuss the modulus stabilization 
by the Casimir energy in the gauge-Higgs unification scenario. 

However, the authors of Ref.~\cite{Garriga:2002vf} show that the KK tower of 
a massless gauge boson provides a negative contribution to 
the radion mass squared. 
Since a contribution of the gluon KK tower is enhanced by the color factor, 
the radion tends to be tachyonic and the extra dimension be destabilized. 
The authors of Ref.~\cite{Garriga:2002vf} also pointed out that 
a non-tachyonic radion mass can be realized by introducing 
gauge kinetic terms localized on the IR brane. 
On the other hand, it is also known that such brane kinetic terms 
affect relations among 
various coupling constants and the KK mass spectra 
in four-dimensional (4D) effective theory. 
Hence it is expected that the magnitudes of the brane kinetic terms 
receive some constraints from the current experimental results. 
It is nontrivial whether the modulus is stabilized or not 
within the allowed region of the parameter space of a model. 

The purpose of this paper is to discuss the modulus stabilization 
by the Casimir energy in a specific gauge-Higgs unification model 
in the warped spacetime, 
including the IR-brane kinetic terms for the gauge fields. 
We also investigate effects of the brane kinetic terms 
on the 4D coupling constants and the first KK gluon mass 
to obtain constraints on the magnitudes of the brane kinetic terms. 

The paper is organized as follows. 
In the next section, we provide a brief review of $SO(5)\times U(1)_X$ 
gauge-Higgs unification model, including the IR-brane kinetic terms 
for the gauge fields. 
In Sec.~\ref{V_eff}, the one-loop effective potential for the radion 
and the Higgs field is shown. 
In Sec.~\ref{Rad_stb}, we calculate the masses of 
the radion and the Higgs boson in the presence of the brane kinetic terms. 
In Sec.~\ref{IR-brane}, we discuss effects of the IR-brane kinetic terms 
on the electroweak gauge couplings of fermions,  
and the mass of the first KK gluon. 
Sec.~\ref{summary} is devoted to the summary and discussions. 
We collect functions that determine the mass spectrum in each sector 
in Appendix~\ref{spectrum}, and provide a brief derivation of 
the one-loop effective potential in Appendix~\ref{one-loop_V}.

\section{$\bdm{SO(5)\times U(1)_X}$ model}
In this section, we briefly review the $SO(5)\times U(1)_X$ gauge-Higgs 
unification model, which was first discussed in Ref.~\cite{Agashe:2004rs}. 
Several similar models with different matter sectors have been studied so far. 
Here we consider a model proposed in Ref.~\cite{Hosotani:2008tx} 
as an example. 

We consider the 5D gauge theory compactified on an orbifold~$S^1/Z_2$. 
The background metric is given by 
\be
 ds^2 = G_{MN}dx^M dx^N = e^{-2\sgm(y)}\eta_{\mu\nu}dx^\mu dx^\nu+dy^2, 
\ee
where $M,N=0,1,2,3,4$ are 5D indices and $\eta_{\mu\nu}=\diag(-1,1,1,1)$. 
The fundamental region of $S^1/Z_2$ is $0\leq y\leq L$. 
The function~$e^{\sgm(y)}$ is a warp factor, and 
$\sgm(y)=ky$ in the fundamental region, where $k$ is 
the inverse AdS curvature radius. 
The orbifold has two fixed points~$y=0$ and $y=L$, which are called 
the UV and IR branes, respectively. 

\subsection{Bulk Lagrangian}
The model has gauge fields~$A^{(G)}_M$, $A_M$ and $B_M$ 
for $SU(3)_C$, $SO(5)$ and $U(1)_X$, respectively. 
In this article, we consider 5D fermions~$\Psi_i$ ($i=1,2,\cdots$) belonging to 
the vectorial representation of $SO(5)$ as matter fields. 
The 5D bulk Lagrangian is given by 
\bea
 \cL \eql \sqrt{-G}\left[-\frac{1}{4}\tr\brc{F_{MN}^{(G)}F^{(G)MN}} 
 -\frac{1}{4}\tr\brc{F_{MN}^{(A)}F^{(A)MN}}-\frac{1}{4}F_{MN}^{(B)}F^{(B)MN}
 \right. \nonumber\\
 &&\hspace{10mm}\left. 
 +\sum_i\brc{
 i\bar{\Psi}_i\Gm^N\cD_N\Psi_i-iM_{\Psi i}\vep(y)\bar{\Psi}_i\Psi_i}\right]
 +\cL_{\rm bd}+\cdots,  \label{5D_L}
\eea
where $G\equiv\det(G_{MN})$, $\Gm^N$ are 5D gamma matrices contracted by 
the f\"{u}nfbein, 
$F_{MN}^{(G)}$, $F_{MN}^{(A)}$ and 
$F_{MN}^{(B)}$ are field strengths for the $SU(3)_C$, $SO(5)$ and $U(1)_X$ 
gauge fields, respectively. 
The covariant derivative of $\Psi_i$ is defined as 
\be
 \cD_N\Psi_i \equiv \brkt{\der_N-\frac{1}{4}\omg_N^{\;AB}\Gm_{AB}
 -ig_CA^{(G)}_N-ig_AA_N-ig_BQ_XB_N}\Psi_i, 
\ee
where $\omg_N^{AB}$ are the spin connection, 
$\Gm^{AB}\equiv\frac{1}{2}\sbk{\Gm^A,\Gm^B}$, 
and $g_C$, $g_A$ and $g_B$ are 
5D gauge coupling constants for $SU(3)_C$, $SO(5)$ and $U(1)_X$, respectively. 
The bulk mass parameters of the fermions~$M_{\Psi i}$ are associated with 
a periodic step function~$\vep(y)$, which is required 
in order for the mass terms to be invariant under the orbifold parity. 
Terms denoted as $\cL_{\rm bd}$ represent brane-localized terms. 
The ellipsis in (\ref{5D_L}) denotes the gauge-fixing terms and 
the ghost terms. 

The orbifold boundary conditions at $y_0\equiv 0$ and $y_1\equiv L$ are 
given by 
\bea
 &&A_\mu^{(G)}(x,y_j-y) = A_\mu^{(G)}(x,y_j+y), \nonumber\\
 &&A_\mu(x,y_j-y) = P_j A_\mu(x,y_j+y)P_j^{-1}, \nonumber\\
 &&B_\mu(x,y_j-y) = B_\mu(x,y_j+y), \nonumber\\
 &&\Psi_i(x,y_j-y) = P_j\Gm^5\Psi_i(x,y_j+y), \nonumber\\
 &&P_j = \diag(-1,-1,-1,-1,+1), \;\;\;\;\; (j=0,1), 
 \label{Z2_parity}
\eea
which reduce the $SU(3)_C\times SO(5)\times U(1)_X$ symmetry to 
$SU(3)_C\times SO(4)\times U(1)_X$. 
The orbifold parities for $A_y^{(G)}$, $A_y$, $B_y$ are 
opposite to those for $A_\mu^{(G)}$, $A_\mu$, $B_\mu$. 

\subsection{Boundary terms}
The boundary conditions in (\ref{Z2_parity}) can be changed by introducing 
4D scalar fields localized on the boundaries whose VEVs give 
brane-localized masses to the 5D fields.\footnote{
The introduction of elementary scalar fields is not essential. 
The boundary conditions can also be changed by condensate of 
fermion bilinear through some strong dynamics. 
} 
Here we introduce a scalar field~$\Phi(x)$ on the UV brane 
which belongs to $(0,\frac{1}{2})$ representation of 
$SO(4)\sim \suL\times\suR$ and has a charge of $U(1)_X$. 
Then the $\suR\times U(1)_X$ symmetry breaks down to $U(1)_Y$, 
similar to the Higgs mechanism in the standard model. 
As a result, $A_\mu^{1_{\rm R}}$, 
$A_\mu^{2_{\rm R}}$ and $A_\mu^{\prime 3_{\rm R}}$ acquire large masses 
at the UV brane. 
Here
\bea
 &&\begin{pmatrix} A_M^{\prime 3_{\rm R}} \\ A_M^Y \end{pmatrix} 
 = \begin{pmatrix} \cph & -\sph \\ \sph & \cph \end{pmatrix}
 \begin{pmatrix} A_M^{3_{\rm R}} \\ B_M \end{pmatrix}, \nonumber\\
 &&\cph \equiv \frac{g_A}{\sqrt{g_A^2+g_B^2}}, \;\;\;\;\;
 \sph \equiv \frac{g_B}{\sqrt{g_A^2+g_B^2}}. 
\eea
Since the typical energy scale at the UV brane is the Planck scale, 
it is natural to assume that the VEV of $\Phi$ is much larger than 
the KK mass scale~$\mKK$. 
Then the net effect for low-lying modes is that they effectively 
obey Dirichlet boundary conditions at the UV brane. 
Other effects of the introduction of $\Phi$ are irrelevant to 
the physics below $\mKK$. 


It is useful to express the $SO(5)$ vector~$\Psi=(\psi_1,\cdots,\psi_5)^t$ as 
\be
 \Psi = \sbk{\begin{pmatrix} \hat{\psi}_{11} \\ \hat{\psi}_{21} 
 \end{pmatrix}, \begin{pmatrix} \hat{\psi}_{12} \\ \hat{\psi}_{22} 
 \end{pmatrix}, \psi_5}, 
\ee
where 
\be
 \hat{\psi} = \begin{pmatrix} \hat{\psi}_{11} & \hat{\psi}_{12} \\
 \hat{\psi}_{21} & \hat{\psi}_{22} \end{pmatrix} 
 \equiv \frac{1}{\sqrt{2}}\brkt{\psi_4+i\vec{\psi}\cdot\vec{\sgm}}i\sgm_2  
\ee
is a bidoublet for $\suL\times\suR$, and 
$\psi_5$ is a singlet under $\suL\times\suR$. 
Then the quarks in the third generation, for instance, are composed of 
two 5D Dirac fermions 
\bea
 \Psi_1 \eql \sbk{Q_1=\begin{pmatrix} T \\ B \end{pmatrix}, 
 q=\begin{pmatrix} t \\ b \end{pmatrix}, t'}, \nonumber\\
 \Psi_2 \eql \sbk{Q_2=\begin{pmatrix} U \\ D \end{pmatrix}, 
 Q_3=\begin{pmatrix} X \\ Y \end{pmatrix}, b'}, 
\eea
and 4D right-handed fermions localized on the UV brane, 
which belong to the $(\frac{1}{2},0)$ representation in $\suL\times\suR$, 
\be
 \hat{\chi}_{1R} = \begin{pmatrix} \hat{T}_R \\ \hat{B}_R \end{pmatrix}, 
 \;\;\;\;\;
 \hat{\chi}_{2R} = \begin{pmatrix} \hat{U}_R \\ \hat{D}_R \end{pmatrix}, 
 \;\;\;\;\;
 \hat{\chi}_{3R} = \begin{pmatrix} \hat{X}_R \\ \hat{Y}_R \end{pmatrix}. 
\ee
The $U(1)_X$ charges of $\Psi_1$, $\Psi_2$, $\hat{\chi}_{1R}$, 
$\hat{\chi}_{2R}$ and $\hat{\chi}_{3R}$ are $2/3$, $-1/3$, $7/6$, $1/6$ 
and $-5/6$, respectively.\footnote{ 
The resulting $\uy$ and $\uem$ charges of each component are listed 
in Sec.~2 of Ref.~\cite{Hosotani:2008tx} or 
in Table~2 in Ref.~\cite{Hosotani:2009qf}. 
}

The symmetry breaking by $\Phi(x)$ on the UV brane can also induce 
the following fermion mass terms localized there. 
\bea
 \cL_{\rm bd}^{\rm fermion} \eql 2i\sqrt{-g}\left\{
 \sum_{\alp=1}^3\bar{\hat{\chi}}_{\alp R}\bar{\sgm}^\mu\cD_\mu\chi_{\alp R}
 -\sum_{\alp=1}^3\mu_\alp\brkt{\bar{\hat{\chi}}_{\alp R}Q_{\alp L}
 -\bar{Q}_{\alp L}\hat{\chi}_{\alp R}} \right. \nonumber\\
 &&\hspace{15mm}\left. 
 -\tl{\mu}\brkt{\bar{\hat{\chi}}_{2R} q_L-\bar{q}_L\hat{\chi}_{2R}}
 \right\}\dlt(y), 
\eea
where $\sqrt{-g}\equiv\det(g_{\mu\nu})$, $g_{\mu\nu}$ is 
the 4D induced metric on the UV brane. 
The brane mass parameters~$\mu_\alp$ ($\alp=1,2,3$) and $\tl{\mu}$ have 
mass dimensions 1/2. 
In the subsequent discussions, 
we suppose that they are much larger than $\mKK$. 
Then the ratio~$\tl{\mu}/\mu_2$ becomes the only relevant parameter 
for physics below $\mKK$. 
In this paper, we neglect the flavor-mixings in quark and lepton sectors 
for simplicity. 
They can always be incorporated by promoting the brane mass parameters 
to matrices. 

Besides the above brane-localized mass terms, 
there can be brane-localized kinetic terms.\footnote{
Such terms will be generically induced by quantum loop effects 
of the bulk fields~\cite{Georgi:2000ks}. 
} 
As we will mention in the next section, the gauge kinetic terms localized 
on the IR brane are necessary to stabilize the radion. 
Thus we introduce the following terms on the IR brane. 
\be
 \cL_{\rm bd}^{\rm kin} = 2\sqrt{-g}\sbk{
 -\frac{\kp_c}{4k}\tr\brc{F_{\mu\nu}^{(G)}F^{(G)\mu\nu}}
 -\frac{\kp_w}{4k}\tr\brc{F_{\mu\nu}^{(A)}F^{(A)\mu\nu}}
 -\frac{\kp_x}{4k}F_{\mu\nu}^{(B)}F^{(B)\mu\nu}}\dlt(y-L), 
\ee
where $\kp_c$, $\kp_w$, $\kp_x$ are dimensionless constants. 
For simplicity, we do not consider kinetic terms on the UV brane 
or brane kinetic terms for the 5D fermions in this paper.

\subsection{Mass spectrum}
Now we calculate the mass spectrum~$\brc{m_n}$ in the 4D effective theory. 
It is determined as solutions to the equation, 
\be
 \rho_I(\lmd_n) = 0, \label{mass_det}
\ee
where $I=G,W,{\rm nt},5/3,2/3,-1/3,-4/3$ specifies the sectors, and 
$\lmd_n\equiv m_n/k$. 
The functions~$\rho_I(\lmd)$ are listed in Appendix~\ref{spectrum}. 

For example, the W and Z boson masses are obtained as the smallest solution 
to $\rho_W(\lmd_W)=0$ and the second smallest solution 
to $\rho_{\rm nt}(\lmd_Z)=0$,\footnote{
The smallest solution to $\rho_{\rm nt}(\lmd)=0$ is $\lmd=0$, which 
corresponds to the massless photon. } 
and are approximately expressed as 
\bea
 m_W \eql k\lmd_W \simeq \frac{ke^{-kL}\sin\thH}{\sqrt{kL+\kp_w}}, 
 \nonumber\\
 m_Z \eql k\lmd_Z \simeq \sqrt{\frac{2\sph^2(kL+\kp_w)+\cph^2(kL+\kp_x)}
 {\sph^2(kL+\kp_w)+\cph^2(kL+\kp_x)}}\frac{ke^{-kL}\sin\thH}{\sqrt{kL+\kp_w}}. 
 \label{ap:mWZ}
\eea

The masses of the top and bottom quarks are obtained as the lowest solutions 
to $\rho_{2/3}(\lmd_t)=0$ and $\rho_{-1/3}(\lmd_b)=0$, respectively. 
In the case of $M_{\Psi 1}=M_{\Psi 2}$ which we assume in the following, 
their approximate expressions are simplified as 
\bea
 m_t \eql k\lmd_t \simeq \frac{k\sqrt{1-4c_t^2}\sin\thH}{\sqrt{2}e^{kL}} 
 \simeq \sqrt{\frac{(1-4c_t^2)(kL+\kp_w)}{2}}m_W, 
 \nonumber\\
 m_b \eql k\lmd_b \simeq \frac{\tl{\mu}}{\mu_2}m_t, 
\eea
where $c_t\equiv M_{\Psi 1}/k=M_{\Psi 2}/k$. 
The above expressions are valid when $c_t<1/2$. 
As we will see in the next section, the effective potential determines 
$\thH=\pi/2$. 
Then the realistic top quark mass is obtained by choosing $c_t\simeq 0.43$ 
for $e^{kL}=10^{15}$ and $\kp_w\ll kL$.

\section{Radion-Higgs potential} \label{V_eff}
The one-loop effective potential for the radion and the Higgs field 
is calculated from the formula~(\ref{expr:V}) with (\ref{def:tauveff}) 
in Appendix~\ref{one-loop_V} obtained 
by the technique in Ref.~\cite{Garriga:2000jb}. 
Noticing that 
$\ln\brc{1-e^{i(\bt-\gm)\pi}\frac{I_\bt(we^{-kL})K_\gm^\kp(w)}
{K_\bt(we^{-kL})I_\gm^\kp(w)}}$ 
is exponentially small\footnote{
The functions~$I^\kp_\gm(u)$ and $K^\kp_\gm(u)$ are 
defined in (\ref{def:IK^kp}).
} for $w\simlt\cO(1)$ unless $\bt\simeq 0$, 
only the gauge fields ($\bt=0$) and the top and bottom quark multiplets 
($\bt=c_t-\frac{1}{2}\simeq -0.03$) can contribute 
to the effective potential~$V$~\cite{Garriga:2002vf}. 
In other words, only the modes whose mode functions spread over the bulk 
can give sizable contributions to $V$. 
In fact, the contribution of the graviton KK tower is exponentially suppressed 
because the graviton is localized around the UV brane 
and $\bt=1$~\cite{Garriga:2000jb}. 
Here the orders of the Bessel functions~$\bt$ and $\gm$ are determined by 
the boundary conditions at the UV and IR branes, respectively. 
If we introduce the UV-brane kinetic terms, they effectively shift $\bt$ 
from zero and make the gauge field contributions negligible. 
So we do not consider the UV-brane kinetic terms in this paper. 
Then the effective potential is expressed as the following form. 
\bea
 V(kL,\thH) \eql \frac{k^4}{16\pi^2}\sbk{
 \tau_{\rm UV}+\tau_{\rm IR}e^{-4kL}
 +e^{-4kL}\int_0^\infty\dr w\;w^3v_{\rm eff}(w;kL,\thH)}, 
\eea
where dimensionless constants~$\tau_{\rm UV}$ and $\tau_{\rm IR}$ 
are associated with tensions at the UV and IR branes, and cannot be 
determined in the context of the 5D field theory. 
The integrand is given from (\ref{def:tauveff}) by 
\bea
 v_{\rm eff}(w;kL,\thH) \eql 24\ln\brc{1-\frac{I_0(we^{-kL})K_0^{\kp_c}(w)}
 {K_0(we^{-kL})I_0^{\kp_c}(w)}}
 +9\ln\brc{1-\frac{I_0(we^{-kL})K_0^{\kp_w}(w)}
 {K_0(we^{-kL})I_0^{\kp_w}(w)}} \nonumber\\
 &&+3\ln\brc{1-c_\phi^2\frac{I_0(we^{-kL})K_0^{\kp_x}(w)}
 {K_0(we^{-kL})I_0^{\kp_x}(w)}
 -s_\phi^2\frac{I_0(we^{-kL})K_0^{\kp_w}(w)}{K_0(we^{-kL})I_0^{\kp_w}(w)}} 
 \nonumber\\
 &&-24\ln\brc{1-\frac{I_{c_t-\frac{1}{2}}(we^{-kL})K_{c_t-\frac{1}{2}}(w)}
 {K_{c_t-\frac{1}{2}}(we^{-kL})I_{c_t-\frac{1}{2}}(w)}} 
 +6\ln\brc{1+\frac{e^{kL}\sin^2\thH}{2w^2\hat{F}_{0,0}^{\kp_w}
 \hat{F}_{1,1}^0}} \nonumber\\
 &&+3\ln\brc{1+\frac{e^{kL}\brkt{
 c_\phi^2\hat{F}_{0,0}^{\kp_x}\hat{F}_{1,0}^{\kp_w}
 +2s_\phi^2\hat{F}_{1,0}^{\kp_x}\hat{F}_{0,0}^{\kp_w}}\sin^2\thH}
 {2w^2\brkt{c_\phi^2\hat{F}_{0,0}^{\kp_x}\hat{F}_{1,0}^{\kp_w}
 +s_\phi^2\hat{F}_{1,0}^{\kp_x}\hat{F}_{0,0}^{\kp_w}}
 \hat{F}_{0,0}^{\kp_w}\hat{F}_{1,1}^0}} \nonumber\\
 &&-12\ln\brc{1+\frac{e^{kL}\sin^2\thH}
 {2w^2\hat{F}_{c_t+\frac{1}{2},c_t+\frac{1}{2}}^0
 \hat{F}_{c_t-\frac{1}{2},c_t-\frac{1}{2}}^0}},  
 \label{expr:veff}
\eea
where the arguments of $\hat{F}_{\alp,\bt}^\kp$ are all $we^{-kL}$. 
We have neglected a small depencence on 
$\abs{\tl{\mu}/\mu_2}^2=(m_b/m_t)^2$, 
and used an approximation~(\ref{ap:rho/KI}). 
The contribution from the region~$w\gg\cO(1)$ is negligible 
in the integral. 
The above $V$ can be understood as the effective potential 
for the radion and the Higgs field 
by promoting the parameters~$kL$ and $\thH$ 
to 4D dynamical fields~$kL(x)$ and $\thH(x)$. 

From the stationary condition for $kL$, we obtain 
\be
 \tau_{\rm IR} = \int_0^\infty\dr w\;w^3\brc{\frac{\der_{kL}v_{\rm eff}}{4}
 -v_{\rm eff}}. \label{stationary_cd}
\ee
By means of this equation, 
we can always choose $\tau_{\rm IR}$ so that the potential has 
a stationary point at a desired value of $kL$. 
In fact, a large warp factor~$e^{kL}=10^{15}$ is realized by 
an $\cO(1)$ value of $\tau_{\rm IR}$. 

From (\ref{expr:veff}), we can see that 
$\thH=\pi/2$ always satisfies the stationary condition for $\thH$. 
As shown in Ref.~\cite{Hosotani:2008tx}, it is a minimum of the potential 
along the $\thH$-direction for relatively large values of the warp factor 
in the absence of the brane kinetic terms. 
This is also true for $\kp_{c,w,x}\neq 0$.

\section{Modulus stabilization} \label{Rad_stb}
Now we consider the stabilization of the size of the extra dimension. 
In this section, we assume a value of $\tau_{\rm IR}$ so that 
$e^{kL}=10^{15}$ is a stationary point of the potential 
along the $kL$-direction. 
Then the AdS curvature scale~$k$ is determined by 
$k\simeq e^{kL}\sqrt{kL+\kp_w}m_W/\sin\thH$, 
which is obtained from (\ref{ap:mWZ}),  
and the typical KK mass scale~$\mKK$ is estimated as 
\be
 \mKK \equiv \frac{\pi k}{e^{kL}-1} 
 \simeq \frac{\pi\sqrt{kL+\kp_w}m_W}{\sin\thH}.  \label{def:mKK}
\ee

The second derivatives of the potential are given as 
\bea
 \der_{kL}^2V \eql \frac{k^4e^{-4kL}}{16\pi^2}
 \int_0^\infty\dr w\;w^3\brc{\der_{kL}^2v_{\rm eff}-4\der_{kL}v_{\rm eff}}, 
 \nonumber\\
 \der_{kL}\der_{\thH}V \eql \frac{k^4e^{-4kL}}{16\pi^2}
 \int_0^\infty\dr w\;w^3\brc{\der_{kL}\der_{\thH}v_{\rm eff}
 -4\der_{\thH}v_{\rm eff}}, \nonumber\\
 \der_{\thH}^2V \eql \frac{k^4e^{-4kL}}{16\pi^2}
 \int_0^\infty\dr w\;w^3\der_{\thH}^2v_{\rm eff}. 
\eea
In the first equation, we have used (\ref{stationary_cd}). 

Note that there is no radion-Higgs mixing in our model because 
$\der_{kL}\der_{\thH}V\propto\cos\thH$ vanishes at $\thH=\pi/2$. 
Thus, the radion mass is calcuated by canonically normalizing 
the radion kinetic term in the Einstein-Hilbert term as 
\bea
 m_{\rm rad}^2 \eql \frac{e^{2kL}-1}{3M_5^3k}\cdot k^2\der_{kL}^2V \nonumber\\
 \sma \frac{k^5e^{-2kL}}{48\pi^2M_5^3}\int_0^\infty\dr w\;w^3
 \brc{\der_{kL}^2v_{\rm eff}-4\der_{kL}v_{\rm eff}}, 
\eea
where $M_5$ is the 5D Planck scale. 
The right-hand sides are evaluated at the minimum 
of the potential~$(e^{kL},\thH)=(10^{15},\pi/2)$. 
Since the 4D Planck scale~$M_{\rm Pl}$ is related to $M_5$ through 
\be
 M_{\rm Pl}^2 \simeq \frac{M_5^3}{2k}, 
\ee
we obtain 
\be
 m_{\rm rad} \simeq \frac{kL+\kp_w}{4\sqrt{6}\pi}\frac{m_W^2e^{kL}}{M_{\rm Pl}}
 \brc{\int_0^\infty\dr w\;w^3
 \brkt{\der_{kL}^2v_{\rm eff}-4\der_{kL}v_{\rm eff}}}^{1/2}. 
\ee

The Higgs mass is calculated as
\bea
 m_H^2 \eql \frac{g_A^2(e^{2kL}-1)}{4k}\der_{\thH}^2V 
 \simeq \frac{g_A^2k^3e^{-2kL}}{64\pi^2}\int_0^\infty\dr w\;w^3
 \der_{\thH}^2v_{\rm eff}. 
\eea
Namely, 
\be
 m_H \simeq \frac{g_4(kL+\kp_w)m_W}{8\pi}
 \brc{\int_0^\infty\dr w\;w^3\der_{\thH}^2v_{\rm eff}}^{1/2}, 
\ee
where 
\be
 g_4 \equiv \frac{g_A\sqrt{k}}{\sqrt{kL+\kp_w}}  \label{expr:g4} 
\ee
is the 4D effective weak gauge coupling.\footnote{
This constant is an approximate expression of the actual gauge coupling 
calculated as an overlap integral of the mode functions. 
} 

\begin{figure}[t]
\centering  \leavevmode
\includegraphics[width=72mm]{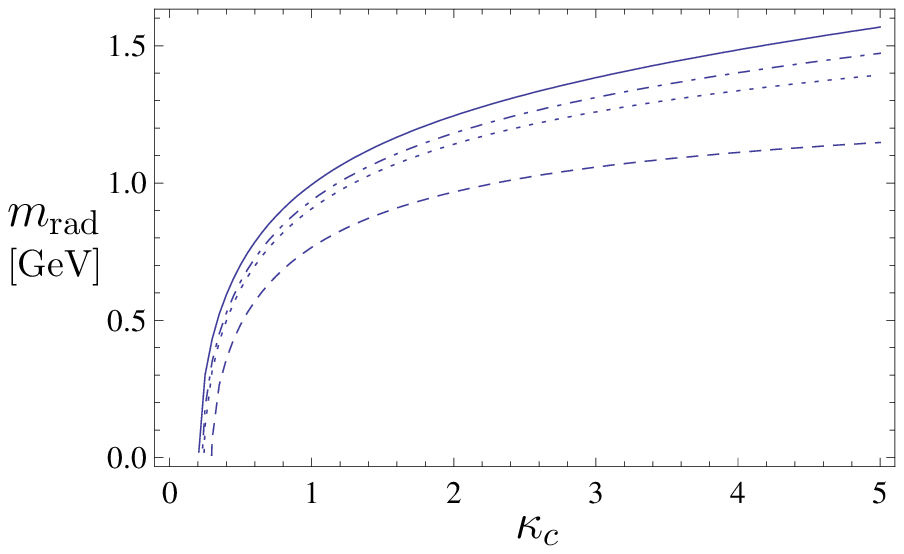} \hspace{10mm}
\includegraphics[width=72mm]{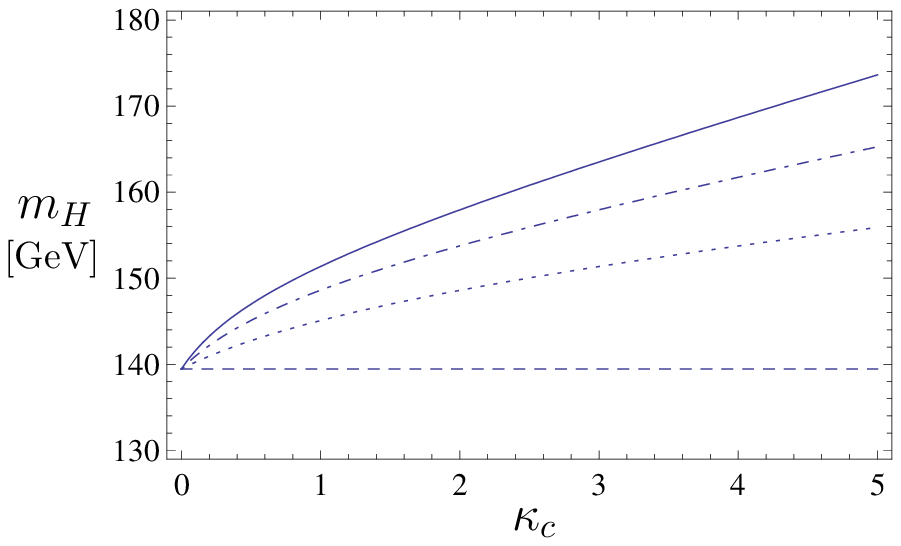}
\caption{The masses of the radion and the Higgs boson as functions of 
the coefficients of the brane kinetic terms. 
The solid, dotdashed, dotted and dashed lines represent 
the case of $(\kp_w,\kp_x)=(1,1)\kp_c, (2/3,1/3)\kp_c, 
(1/3,2/3)\kp_c$ and $(0,0)$, respectively. }
\label{Radion_mass}
\end{figure}
In Fig.~\ref{Radion_mass}, we show the radion mass and the Higgs boson mass 
as functions of the coefficients of the brane kinetic terms~$\kp_{c,w,x}$. 
In the absence of the brane kinetic terms, the radion mass is tachyonic 
and the modulus is not stabilized. 
If we turn on them, the radion mass squared monotonically increases 
as a function of $\kp_{c,w,x}$. 
Due to the color factor, the gluon provides 
the largest contribution to the radion mass. 
In the case of $(\kp_w,\kp_x)=(1,1)\kp_c,(2/3,1/3)\kp_c,(1/3,2/3)\kp_c$ 
and $(0,0)$, the radion becomes non-tachyonic 
for $\kp_c\simgt 0.21,0.24,0.25$ and 0.30, respectively. 
When $\kp_c=0$, 
much larger values of $\kp_{w,x}$ are necessary to stabilize the modulus. 
For example, $\kp_w\simgt 7.2, 8.6,12$ when $\kp_w=\kp_x/2$, 
$\kp_w=\kp_x$, $\kp_w=2\kp_x$, respectively.   
We find that $\cO(1)$ values of $\kp_{c,w,x}$ lead to the radion mass 
around 1~GeV.

\section{Effects of IR-brane kinetic terms} \label{IR-brane}
Although the brane kinetic terms for the gauge fields are necessary 
for the modulus stabilization, large brane kinetic terms can deviate 
the weak boson couplings to the fermions from the standard model values, 
as shown in Ref.~\cite{Davoudiasl:2004pw}. 
This is because they repel the mode functions of the gauge bosons away 
from the IR brane, where the custodial symmetry~$SO(4)$ exists. 
The Weinberg angle~$\thw$ is defined by the ratio of the W and Z bosons as 
\bea
 \sin^2\thw \defa 1-\frac{m_W^2}{m_Z^2} 
 \simeq \frac{\sph^2(kL+\kp_w)}{2\sph^2(kL+\kp_w)+\cph^2(kL+\kp_x)}. 
 \label{def:thw_os}
\eea
We have used (\ref{ap:mWZ}). 
The value of $\sph$, or $g_A/g_B$, is determined for given values of 
$\kp_w$ and $\kp_x$ so that the above defined Weinberg angle takes 
the correct value~$\sin^2\thw\simeq 0.22$. 

On the other hand, the Weinberg angle is also defined by the ratio 
of the gauge couplings of the fermions to the photon and the W boson. 
For example, let us consider the gauge couplings of the quarks 
in the first generation. 
Then 
\bea
 \cL_{\rm gauge}^{(4)} \eql eA_\mu^{\gm(0)}\brc{
 \frac{2}{3}\brkt{\bar{u}_L\gm^\mu u_L+\bar{u}_R\gm^\mu u_R}
 -\frac{1}{3}\brkt{\bar{d}_L\gm^\mu d_L+\bar{d}_R\gm^\mu d_R}} \nonumber\\
 &&+\frac{g_{ud,L}^{(W)}}{\sqrt{2}}\brkt{W_\mu\bar{d}_L\gm^\mu u_L+\hc}
 +\frac{g_{ud,R}^{(W)}}{\sqrt{2}}\brkt{W_\mu\bar{d}_R\gm^\mu u_R+\hc}
 \nonumber\\
 &&+\frac{1}{\cos\thw}Z_\mu\brc{g_{uL}^{(Z)}\bar{u}_L\gm^\mu u_L
 +g_{uR}^{(Z)}\bar{u}_R\gm^\mu u_R
 +g_{dL}^{(Z)}\bar{d}_L\gm^\mu d_L+g_{dR}^{(Z)}\bar{d}_R\gm^\mu d_R} 
 \nonumber\\
 &&+\cdots. 
\eea
Each coupling constant is given as overlap integral 
of the relevant mode functions. 
For example, the electromagnetic coupling constant~$e$ is calculated as
\be
 e = \frac{g_A\sqrt{k}\sph}{\sqrt{2\sph^2(kL+\kp_w)+\cph^2(kL+\kp_x)}} 
 \simeq \frac{g_A\sqrt{k}}{\sqrt{kL+\kp_w}}\sin\thw. 
 \label{expr:e}
\ee
The absolute value of the 5D coupling~$g_A$ is fixed for given values of 
$\kp_w$ and $\kp_x$ so that $e$ takes the observed value. 
We have used (\ref{def:thw_os}) in the second equality. 
From this, we can read off the approximate expression of 
the weak gauge coupling~$g_4$ shown in (\ref{expr:g4}). 
We do not show the explicit forms of other coupling constants here, 
but they are obtained from those given in Ref.~\cite{Hosotani:2009qf}
by modifying the mode functions of the gauge bosons 
including the brane kinetic terms. 
In contrast to the standard model, the W boson couplings of 
the right-handed quarks do not completely vanish although 
they are negligibly small. 
Then the Weinberg angle is defined, for example, by 
\be
 \sin\bthw \equiv \frac{e}{g_{ud,L}^{(W)}}. 
\ee

\begin{figure}[t]
\centering  \leavevmode
\includegraphics[width=70mm]{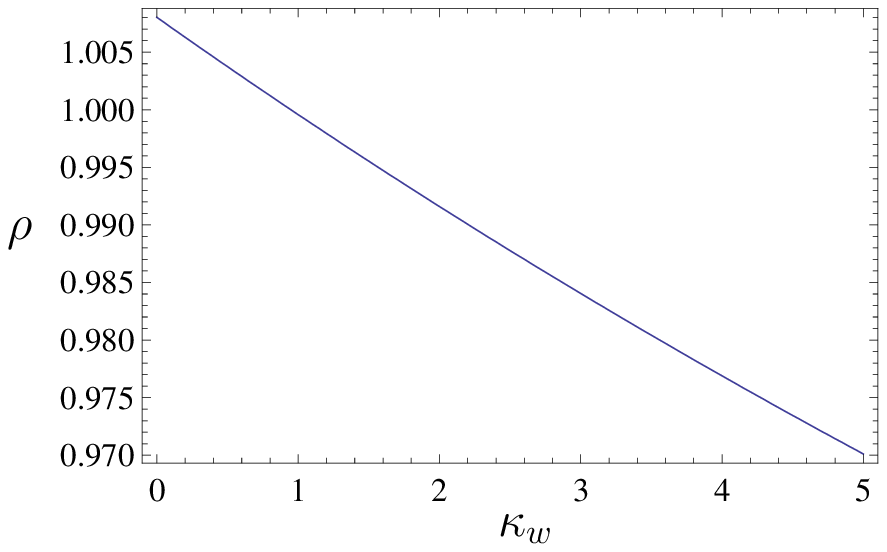} \hspace{10mm}
\includegraphics[width=70mm]{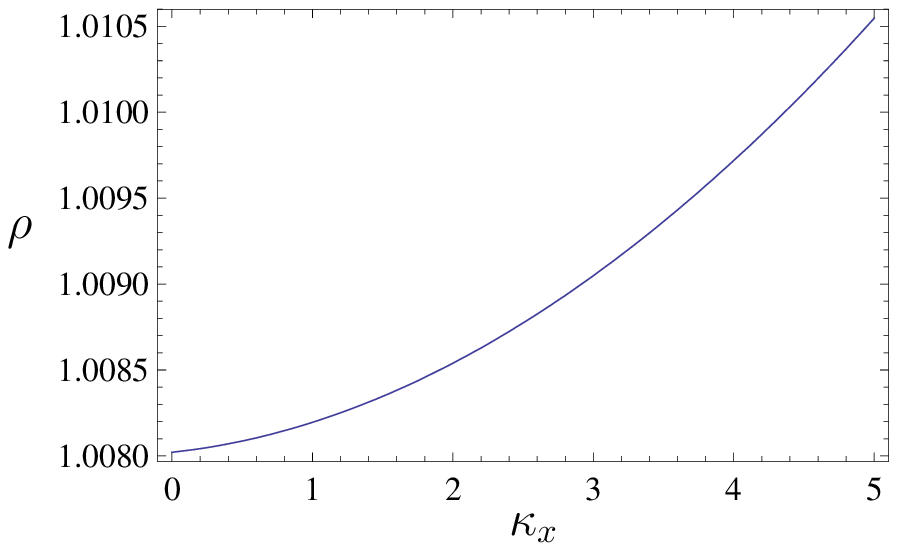}
\caption{The $\rho$ parameter defined in (\ref{def:rho}) as a function of 
$\kp_w$ when $\kp_x=0$ (left figure) and 
of $\kp_x$ when $\kp_w=0$ (right figure). }
\label{sin_thw}
\end{figure}
By utilizing (\ref{def:thw_os}) and (\ref{expr:e}), 
we can estimate the $\rho$ parameter defined by 
\be
 \rho \equiv \frac{m_W^2}{m_Z^2\cos^2\bthw}
 = \frac{\cos^2\thw}{\cos^2\bthw}.  \label{def:rho}
\ee
In Fig.~\ref{sin_thw}, we show this 
as functions of $\kp_w$ and $\kp_x$. 
Current electroweak data fitting favors the $\rho$ parameter being close 
to one, \ie, $1.00989 \leq \rho^{\rm exp} \leq 1.01026$~\cite{Yao:2006px}. 
Similar to the result in Ref.~\cite{Davoudiasl:2004pw}, the brane kinetic 
terms for $SO(4)\subset SO(5)$ and $U(1)_X$ deviate the Weinberg angle 
in the opposite directions. 
The former reduces the value of $\rho$ (and thus $\bthw$) 
while the latter raises them. 
Therefore there is a parameter region where $\rho$ stays within 
the experimental error even for large $\kp_w$ and $\kp_x$.\footnote{
We have to include the loop contributions when $\rho$ is compared 
with $\rho^{\rm exp}$. } 

Since the mode functions of the W and Z bosons are no longer constants 
after the electroweak symmetry breaking occurs, 
the universality of the gauge couplings to them is generically violated. 
As pointed out in Ref.~\cite{Hosotani:2009qf}, however, 
such violation remains less than 1\% except for the top quark due to 
the left-right symmetry the model has. 
This is true even in the presence of the brane kinetic terms. 
The largest violation appears in the Z boson coupling of the top quark. 
For instance, in the case of $(\kp_w,\kp_x)=(5,0),(5,5),(0,5)$, 
it deviates from the Z boson coupling of the up quark 
by 7-8\% for the left-handed component 
and 16-18\% for the right-handed component. 
The universality violation for the first two generations are less than 
percent level. 

As mentioned in the previous section, the brane kinetic term for 
the gluon is necessary to stabilize the modulus. 
On the other hand, it is well known that such a term lowers 
the first KK gluon mass~$m_{g1}$ compared to the typical 
KK scale~$\mKK$~\cite{Davoudiasl:2004pw,Lillie:2007ve}. 
\begin{figure}[t]
\centering  \leavevmode
\includegraphics[width=70mm]{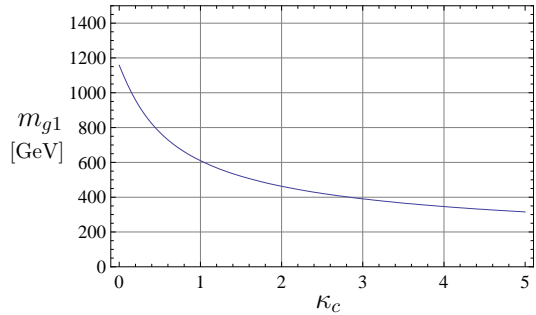} 
\caption{The mass of the first KK gluon~$m_{g1}$ as a function of $\kp_c$. }
\label{KKgluon_mass}
\end{figure}
Fig.~\ref{KKgluon_mass} shows $m_{g1}$ as a function of $\kp_c$. 
The first KK gluon becomes lighter than 600~GeV for $\kp_c>1$. 
Such a light colored particle is problematic and contradicts 
the results of Tevatron searches for resonant~$t\bar{t}$ production. 
Since the experimental lower bound on $m_{g1}$ is 800-900~GeV 
according to Refs.~\cite{Lillie:2007ve,Guchait:2007ux},\footnote{
The precise bound depends on a detail of a model under consideration. } 
a possible maximal value of $\kp_c$ is bounded by around 0.5 
from Fig.~\ref{KKgluon_mass}. 
Recalling that $\kp_c\simgt 0.2-0.3$ is required in order to 
stabilize the modulus, the light KK gluon appears below 1~TeV, 
which marginally satisfies the experimental bound. 

The situation will be improved if we consider models in which  
smaller values of $\thH$ are realized by changing the matter sector. 
Such a model is proposed in Ref.~\cite{Medina:2007hz}, for example. 
Then, the KK mass scale raises by a factor of $1/\sin\thH$ 
(see (\ref{def:mKK})), 
and the deviation of each gauge coupling from the standard model 
becomes smaller than in our model. 
On the other hand, the modulus stabilization is expected to occur 
for $\cO(1)$ values of $\kp_c$ 
because a largest contribution to $m_{\rm rad}^2$ comes from 
the gluon KK tower, which is independent of the change of the matter sector, 
unless we consider a model with a large number of exotic fermion fields.

\section{Summary and discussions} \label{summary}
We have discussed the modulus stabilization of $S^1/Z_2$ 
by the Casimir energy and effects of IR-brane kinetic terms 
for the gauge fields 
in the context of gauge-Higgs unification in the warped spacetime. 
In the absence of the brane kinetic terms, the modulus 
is not stabilized due to a large negative contribution of the gluon loop 
to the radion mass squared~$m_{\rm rad}^2$. 
This can be cured by introducing the brane kinetic terms for the gauge fields 
at the IR brane. 
Especially the brane kinetic term for the gluon provides 
a sizable positive contribution to $m_{\rm rad}^2$ due to a large color factor. 
The modulus is actually stabilized for $\kp_c\simgt 0.2$-$0.3$, and 
the radion obtains a mass of $\cO(\mbox{1~GeV})$ 
for $\cO(1)$ values of $\kp_c$. 
In the case of $\kp_c=0$, much larger values of $\kp_{w,x}$ are necessary 
for the modulus stabilization. 
For example, $\kp_w\simgt 7.2,8.6,12$ is needed 
when $\kp_w=\kp_x/2$, $\kp_w=\kp_x$, $\kp_w=2\kp_x$, respectively. 

As for the modulus stabilization, we can also cancel the large negative 
contribution of the gluon to $m_{\rm rad}^2$ by introducing extra 
colored fermions that belong to the $SU(3)_C$ adjoint representation 
and have boundary conditions such that they do not have zero-modes. 
However, in order for them to give a sizable contribution 
to $m_{\rm rad}^2$, their mode functions 
must obey the Neumann boundary conditions at the UV brane, 
and thus their lightest modes can be heavy at most 400~GeV. 
Such light colored particles are already excluded by the experiments. 

The IR-brane kinetic terms also affect 
4D coupling constants among light modes and the lightest KK masses. 
The brane kinetic term for $SO(4)$ reduces the $\rho$ parameter 
and that for $U(1)_X$ raises it. 
Thus the $\rho$ paremeter can remain within the experimental error 
even for a large value of $\kp_w$ if $\kp_x$ takes an appropriate value. 
The universality violation of the gauge couplings to the W and Z bosons 
are tiny except for the top quark due to the left-right 
symmetry~$\suL\times\suR$, which is checked in Ref.~\cite{Hosotani:2009qf} 
in the case of no brane kinetic terms. 
We have checked that this is also true in the presence of 
the IR-brane kinetic terms for the gauge fields. 
The masses of the lightest KK gauge bosons 
monotonically decrease as $\kp_{c,w,x}$ increase. 
In fact, in the parameter region that ensures the modulus stabilization, 
the KK gluon with the mass~$m_{g1}\simlt 1$~TeV appears,  
which marginally satisfies the experimental 
bound~$m_{g1}\simgt 800-900$~GeV~\cite{Lillie:2007ve,Guchait:2007ux}. 
%

The allowed parameter region will be enlarged if we consider models in which 
smaller values of $\thH$ are realized, 
just like a model in Ref.~\cite{Medina:2007hz}. 
In such models, the KK modes become heavier and the deviation of 
each gauge coupling from the standard model is smaller than in our model, 
while the modulus is stabilized for $\cO(1)$ values of $\kp_c$. 
We should also note that the radion is mixed with the Higgs boson 
when $\thH\neq\pi/2$. 
In addition, the IR-brane kinetic terms also affect 
the violation scale of the tree-level unitarity. 
One of the authors showed that, in the absence of the brane kinetic terms, 
the tree-level unitarity will be violated around the KK mass scale 
in a gauge-Higgs unification model 
in the warped spacetime, irrespective of the values of 
$\thH$~\cite{Haba:2009ei}. 
This means that the perturbative calculation will no longer be reliable 
when the KK modes start to propagate. 
Inclusion of the IR-brane kinetic terms may delay the unitarity violation 
to higher energy scales, 
just like in the 5D Higgsless model~\cite{Davoudiasl:2004pw}. 
We will discuss these issues in a subsequent paper.

\vskip .5cm

\leftline{\bf Acknowledgments}
The authors would like to thank R.~Kitano for a useful discussion.

\appendix

\section{Mass spectrum} \label{spectrum}
Here we collect the expressions of the functions~$\rho_I(\lmd)$ 
that determine the mass spectrum by (\ref{mass_det}) in our model. 
These functions are written in terms of functions defined by
\bea
 F_{\alp,\bt}^\kp(\lmd) \defa 
 J_\alp(\lmd)Y_\bt^\kp(\lmd z_L)-Y_\alp(\lmd)J_\bt^\kp(\lmd z_L), 
\eea
where $z_L\equiv e^{kL}$, and 
\be
 J_\bt^\kp(u) \equiv J_\bt(u)-\kp u J_{\bt+1}(u), \;\;\;\;\;
 Y_\bt^\kp(u) \equiv Y_\bt(u)-\kp u Y_{\bt+1}(u). 
\ee
For a calculation of the effective potential, we also define 
\be
 \hat{F}_{\alp,\bt}^\kp(w) \equiv I_\alp(w)K_\bt^\kp(w z_L)
 -e^{-i(\alp-\bt)\pi}K_\alp(w)I_\bt^\kp(w z_L), 
\ee
where
\be
 I_\bt^\kp(u) \equiv I_\bt(u)+\kp u I_{\bt+1}(u), \;\;\;\;\;
 K_\bt^\kp(u) \equiv K_\bt(u)-\kp u K_{\bt+1}(u).  \label{def:IK^kp}
\ee
Then the following relation holds. 
\be
 F_{\alp,\bt}^\kp(iw) = -\frac{2}{\pi}e^{i(\alp-\bt)\pi/2}
 \hat{F}_{\alp,\bt}^\kp(w). 
\ee
The asymptotic behavior of $\hat{F}^\kp_{\alp,\bt}(w)$ 
for $\Re w\gg 1$ is
\be
 \hat{F}^\kp_{\alp,\bt}(w) = -\frac{e^{w(z_L-1)}}{2w\sqrt{z_L}}
 e^{i(\bt-\alp)\pi}\brkt{1+\kp wz_L}\brc{1+\cO\brkt{w^{-1}}}. 
 \label{asp_hF}
\ee


For the gauge bosons, $\rho_I(\lmd)$ are given by 
\begin{description}
 \item{\bf Gluon sector}
 \be
  \rho_G(\lmd) = \lmd F_{0,0}^{\kp_c}(\lmd). 
 \ee

 \item{\bf W boson sector}
 \be
  \rho_W(\lmd) = F^{\kp_w}_{1,0}(\lmd)\brkt{F_{0,0}^{\kp_w}(\lmd)
 F_{1,1}^0(\lmd)-\frac{2\sin^2\thH}{\pi^2\lmd^2z_L}}.  
 \ee

 \item{\bf Neutral sector}
 \bea
 \rho_{\rm nt}(\lmd) \eql \cph^2\lmd F_{0,0}^{\kp_x}(\lmd)
 F_{1,0}^{\kp_w}(\lmd)\brkt{F_{0,0}^{\kp_w}(\lmd)F_{1,1}^0(\lmd)
 -\frac{2\sin^2\thH}{\pi^2\lmd^2z_L}} 
 \nonumber\\
 &&+\sph^2\lmd F_{1,0}^{\kp_x}(\lmd)F_{0,0}^{\kp_w}(\lmd)
 \brkt{F_{0,0}^{\kp_w}(\lmd)F_{1,1}^0(\lmd)
 -\frac{4\sin^2\thH}{\pi^2\lmd^2z_L}}. 
 \eea

 Especially, when $\kp_w=\kp_x$, the function~$\rho_{\rm nt}(\lmd)$ 
 can be factorized as $\rho_\gm(\lmd)\rho_Z(\lmd)$, 
 and the corresponding KK tower is decomposed into the following two sectors. 
 \begin{description}
  \item{\bf Photon sector}
  \be
   \rho_\gm(\lmd) = \lmd F_{0,0}^{\kp_w}(\lmd). 
  \ee

  \item{\bf Z boson sector}
  \be
   \rho_Z(\lmd) = F_{1,0}^{\kp_w}(\lmd)\brkt{
   F_{0,0}^{\kp_w}(\lmd)F_{1,1}^0(\lmd)
   -\frac{2(1+\sph^2)\sin^2\thH}{\pi^2\lmd^2z_L}}. 
  \ee
 \end{description}

 \item{\bf $\bdm{\hat{4}}$-component sector}
 \be
  \rho_{\hat{4}}(\lmd) = F_{1,0}^0(\lmd). 
 \ee
\end{description}

For quarks, there are four sectors according 
to the $\uem$ charge~$Q_{\rm EM}$. 
\begin{description}
 \item{\bf $\bdm{Q_{\rm EM}=\frac{5}{3}}$ sector}
 \be
  \rho_{5/3}(\lmd) = F_{c+\frac{1}{2},c-\frac{1}{2}}^0(\lmd). 
 \ee

 \item{\bf $\bdm{Q_{\rm EM}=\frac{2}{3}}$ sector}
 \be
  \rho_{2/3}(\lmd) = \brkt{F_{c+\frac{1}{2},c-\frac{1}{2}}^0(\lmd)}^2\brkt{
  F_{c+\frac{1}{2},c+\frac{1}{2}}^0(\lmd)
  F_{c-\frac{1}{2},c-\frac{1}{2}}^0(\lmd)
  -\frac{2\sin^2\thH}{(1+r)\pi^2\lmd^2z_L}}, 
 \ee
 where 
 \be
  r \equiv \abs{\frac{\tl{\mu}}{\mu_2}}^2 = \brkt{\frac{m_b}{m_t}}^2. 
 \ee
 
 \item{\bf $\bdm{Q_{\rm EM}=-\frac{1}{3}}$ sector}
 \be
  \rho_{-1/3}(\lmd) = \brkt{F_{c+\frac{1}{2},c-\frac{1}{2}}^0(\lmd)}^2\brkt{
  F_{c+\frac{1}{2},c+\frac{1}{2}}^0(\lmd)F_{c-\frac{1}{2},c-\frac{1}{2}}^0(\lmd)
  -\frac{2r\sin^2\thH}{(1+r)\pi^2\lmd^2z_L}}. 
 \ee

 \item{\bf $\bdm{Q_{\rm EM}=-\frac{4}{3}}$ sector}
 \be
  \rho_{-4/3}(\lmd) = F_{c+\frac{1}{2},c-\frac{1}{2}}^0(\lmd). 
 \ee
\end{description}
Here we have assumed that the two 5-plet fermions in each generation 
have a common bulk mass, \ie, $c\equiv M_{\Psi 1}/k=M_{\Psi 2}/k$, 
and all the brane mass parameters are assumed to be sufficiently large. 

The lepton sector has a similar structure to the quark sector. 
(See Ref.~\cite{Hosotani:2009qf}.)

\section{One-loop effective potential} \label{one-loop_V}
Here we derive the effective potential for the radion and the Higgs field 
at one-loop level. 
By using the dimensional regularization, 
it is calculated as 
\bea
 V \eql \sum_I \frac{(-)^{2\eta_I}N_I}{2}
 \sum_n\int\frac{d^Dp}{(2\pi)^D}\ln\brkt{p^2+m_{In}^2} \nonumber\\
 \eql \sum_I \frac{(-)^{2\eta_I}N_I}{(4\pi)^{D/2}}
 \frac{\pi}{D\Gm(D/2)\sin(\pi D/2)}\sum_n m_{In}^D, 
 \label{int_expr:V}
\eea
where $D=4+\ep$, $\eta_I=0$ $(1/2)$ for bosons (fermions), $N_I$ is 
a number of degrees of freedom for a particle in sector~$I$. 
The KK mass eigenvalues~$m_{In}$ are solutions to 
\be
 \rho_I(\lmd_{In}) = 0, 
\ee
where $\lmd_{In}\equiv m_{In}/k$, and the functions~$\rho_I(u)$ 
are listed in Appendix~\ref{spectrum}. 
These masses depend on $\thH$ and the warp factor~$z_L=e^{kL}$. 

Here let us define a generalized zeta function as 
\be
 \hat{\zt}(D) \equiv \sum_n\lmd_n^D. 
\ee
This is well-defined for $\Re D<-1$. 
Following the technique of Ref.~\cite{Garriga:2000jb}, 
this is analytically continued to the region~$\Re D<1$ 
and can be traded for the following integral. 
\bea
 \hat{\zt}(D) \eql \frac{D}{\pi}\sin\brkt{\frac{\pi D}{2}}
 \int_0^\infty\dr w\;w^{D-1}\ln\frac{\rho_I(iw)}{\rho^{\rm asp}_I(iw)}, 
 \label{zt2}
\eea 
where $\rho^{\rm asp}_I(u)$ is a $\thH$-independent analytic function 
that satisfies 
\be
 \frac{\rho_I(u)}{\rho^{\rm asp}_I(u)} = 1+\cO\brkt{u^{-1}}, 
\ee
for $\Im u\gg 1$. 
For instance, 
\bea
 \rho_W(iw) \eql -\frac{8i}{\pi^3}\hat{F}_{1,0}^{\kp_w}(w)
 \brc{\hat{F}_{0,0}^{\kp_w}(w)\hat{F}_{1,1}^0(w)+\frac{\sin^2\thH}{2w^2z_L}}, 
 \nonumber\\
 \rho_W^{\rm asp}(iw) \eql -\frac{i}{\pi^3}
 \frac{e^{3w(z_L-1)}}{w^3z_L^{3/2}}\brkt{1+\kp_w wz_L}^2. 
\eea
Here we have used (\ref{asp_hF}). 
In general, $\rho^{\rm asp}_I(u)$ can be expressed as  
$f_I^{\rm UV}(u)f_I^{\rm IR}(uz_L)$, where function forms of 
$f_I^{\rm UV}$ and $f_I^{\rm IR}$ are independent of $\thH$ and $z_L$. 

Note that, except for the neutral sector, 
a $\thH$-independent part~$\rho_{0I}(iw)
\equiv\rho_I(iw)|_{\thH=0}$ has a form of a product of 
\bea
 \hat{F}^\kp_{\alp,\bt}(w) \eql e^{-i(\alp-\bt)\pi}K_\alp(w)I_\bt^\kp(wz_L)
 \brc{1-e^{i(\alp-\bt)\pi}\frac{I_\alp(w)K_\bt^\kp(wz_L)}
 {K_\alp(w)I_\bt^\kp(wz_L)}}. 
\eea
Thus we can define $\cK_I(w)$ and $\cI_I(w)$, 
which are products of $e^{-i\alp\pi}K_\alp(w)$ and $e^{i\bt\pi}I_\bt^\kp(w)$ 
respectively, so that $\rho_{0I}(iw)/\cK_I(w)\cI_I(wz_L)$ becomes 
a product of $\brc{1-e^{i(\alp-\bt)\pi}\frac{I_\alp(w)K_\bt^\kp(wz_L)}
{K_\alp(w)I_\bt^\kp(wz_L)}}$. 
For the neutral sector, we define 
$\cK_{\rm nt}\equiv \frac{4w}{\pi^2}\brc{K_0K_1}^2$ 
and $\cI_{\rm nt}\equiv 
\frac{4}{\pi^2}I_0^{\kp_x}\brkt{I_0^{\kp_w}}^2I_1$. 
Then, 
\bea
 \ln\frac{\rho_{\rm nt}(iw)}{\cK_{\rm nt}(w)\cI_{\rm nt}(wz_L)} 
 \eql \ln\frac{\cph^2\hat{F}_{0,0}^{\kp_x}(w)\hat{F}_{1,0}^{\kp_w}(w)
 +\sph^2\hat{F}_{1,0}^{\kp_x}(w)\hat{F}_{0,0}^{\kp_w}(w)}
 {-K_0(w)K_1(w)I_0^{\kp_x}(wz_L)I_0^{\kp_w}(wz_L)} \nonumber\\
 &&+\ln\brc{1-\frac{I_0(w)K_0^{\kp_w}(wz_L)}{K_0(w)I_0^{\kp_w}(wz_L)}}
 +\ln\brc{1-\frac{I_1(w)K_1(wz_L)}{K_1(w)I_1(wz_L)}} 
 \nonumber\\
 \sma \ln\brc{1-\cph^2\frac{I_0(w)K_0^{\kp_x}(wz_L)}
 {K_0(w)I_0^{\kp_x}(wz_L)}-\sph^2\frac{I_0(w)K_0^{\kp_w}(wz_L)}
 {K_0(w)I_0^{\kp_w}(wz_L)}} \nonumber\\
 &&+\ln\brc{1-\frac{I_0(w)K_0^{\kp_w}(wz_L)}{K_0(w)I_0^{\kp_w}(wz_L)}}, 
 \label{ap:rho/KI}
\eea
for $w\simlt\cO(1)$. 

Therefore, (\ref{int_expr:V}) is rewritten as 
\bea
 V \eql \sum_I \frac{(-)^{2\eta_I}N_Ik^D}{(4\pi)^{D/2}\Gm(D/2)}
 \int_0^\infty\dr w\;w^{D-1}\ln\frac{\rho_I(iw)}
 {\rho_I^{\rm asp}(iw)}
 \nonumber\\
 \eql \sum_I \frac{(-)^{2\eta_I}N_Ik^D}
 {(4\pi)^{D/2}\Gm(D/2)}\int_0^\infty\dr w\;w^{D-1}
 \left(\ln\frac{\cK_I(w)\cI_I(wz_L)}{f_I^{\rm UV}(iw)f_I^{\rm IR}(iwz_L)} 
 \right. \nonumber\\
 &&\hspace{60mm} \left. 
 +\ln\frac{\rho_{0I}(iw)}{\cK_I(w)\cI_I(wz_L)}
 +\ln\frac{\rho_I(iw)}{\rho_{0I}(iw)}\right) \nonumber\\
 \eql \frac{k^D}{(4\pi)^{D/2}\Gm(D/2)}\brc{
 \tau_{\rm UV}+\frac{\tau_{\rm IR}}{z_L^D}
 +\frac{1}{z_L^D}\int_0^\infty\dr w\;w^{D-1}
 v_{\rm eff}(w;kL,\thH)},  \label{expr:V}
\eea
where 
\bea
 \tau_{\rm UV} \defa \sum_I(-)^{2\eta_I}N_I\int_0^\infty\dr w\;w^{D-1}
 \ln\frac{\cK_I(w)}{f_I^{\rm UV}(iw)}, \nonumber\\
 \tau_{\rm IR} \defa \sum_I(-)^{2\eta_I}N_I\int_0^\infty\dr w\;w^{D-1}
 \ln\frac{\cI_I(w)}{f_I^{\rm IR}(iw)}, \nonumber\\
 v_{\rm eff}(w;kL,\thH) \defa \sum_I(-)^{2\eta_I}N_I
 \brc{\ln\frac{\rho_{0I}(iw/z_L)}{\cK_I(w/z_L)\cI_I(w)}
 +\ln\frac{\rho_I(iw/z_L)}{\rho_{0I}(iw/z_L)}}.  
 \label{def:tauveff}
\eea
Note that the third term in the brace in (\ref{expr:V}) is finite 
while the others diverge when we set $D=4$.  
The divergent constants~$\tau_{\rm UV}$ and $\tau_{\rm IR}$ can be 
absorbed in the renormalization of the tensions of the UV and IR branes, respectively. 
Only the second term of $v_{\rm eff}(w;kL,\thH)$ in (\ref{def:tauveff}) 
has a $\thH$-dependence, and corresponds to a contribution calculated 
in Ref.~\cite{Falkowski:2006vi}.


\end{document}